\documentclass[a4paper, 12pt, openbib ]{article}
\usepackage{times}
\usepackage{pifont}
\usepackage{amssymb}
\usepackage{amsfonts}
\usepackage{amsmath}
\usepackage{graphicx}
\usepackage{bm}
\usepackage{geometry}
\usepackage{enumerate}

\usepackage{tikz}
\usetikzlibrary{patterns}

\def\tfract#1/#2{{\textstyle{\raise0.8pt\hbox{$\scriptstyle#1$}\over%
\hbox{\lower0.8pt\hbox{$\scriptstyle#2$}}}}}
\def\mezzo{\tfract 1/2 }
\def\terzo{\tfract 1/3}
\def\dueterzi{\tfract 2/3}
\def\quarto{\tfract 1/4 }
\def\trequarti{\tfract 3/4 }

\def\gmezzi{\tfract g/2 } 
\def\gterzi{\tfract g/3 } 
\def\tremezzi{\tfract 3/2 }

\def\sesto{\tfract 1/6 }

\def\radi2k{\tfract 1/{\sqrt {2k}} } 
\def\der{\partial }
\def\downnormalfill{$\,\,\vrule depth4pt width0.4pt
\leaders\vrule depth 0pt height0.4pt\hfill\vrule depth4pt width0.4pt\,\,$}
\def\WT#1{\mathop{\vbox{\ialign{##\crcr\noalign{\kern3pt}
      \downnormalfill\crcr\noalign{\kern0.8pt\nointerlineskip}
      $\hfil\displaystyle{#1}\hfil$\crcr}}}\limits}

\def\be{\begin{equation}}
\def\ee{\end{equation}}
\def\bes{\begin{equation*}}
\def\ees{\end{equation*}}
\def\bea{\begin{eqnarray}}
\def\eea{\end{eqnarray}}
\def\beas{\begin{eqnarray*}}
\def\eeas{\end{eqnarray*}}
\def\ba{\begin{array}{rcl}}
\def\ea{\end{array}}
\def\der{\partial}
\numberwithin{equation}{section}
\def\go{\leavevmode \raise.3ex\hbox{$\scriptscriptstyle \langle\!\langle\!  $}%
~\ignorespaces}

\def\gf{\relax \ifhmode \unskip~\else \leavevmode \fi \raise.3ex\hbox{$\! \scriptscriptstyle\rangle\!\rangle\, $}}
\usepackage{amsmath,amssymb,amsfonts}	
\usepackage{slashed}            
\usepackage{bbm}                
\usepackage{xspace}				

\usepackage{tikz}
\usetikzlibrary{patterns, arrows, shapes}
\usetikzlibrary{trees}
\usetikzlibrary{matrix, arrows} 				
\usetikzlibrary{positioning}				
\usetikzlibrary{calc,through}				
\usetikzlibrary{decorations.pathreplacing}  
\usepackage{pgffor}							

\usetikzlibrary{decorations.pathmorphing}	
\usetikzlibrary{decorations.markings}
\tikzset{
    vector/.style={decorate, decoration={snake}, draw},
	provector/.style={decorate, decoration={snake,amplitude=2.5pt}, draw},
	antivector/.style={decorate, decoration={snake,amplitude=-2.5pt}, draw},
    fermion/.style={draw=black, postaction={decorate},
        decoration={markings,mark=at position .55 with {\arrow[draw=black]{>}}}},
    fermionbar/.style={draw=black, postaction={decorate},
        decoration={markings,mark=at position .55 with {\arrow[draw=black]{<}}}},
    fermionnoarrow/.style={draw=black},
    gluon/.style={decorate, draw=black,
        decoration={coil,amplitude=4pt, segment length=5pt}},
    scalar/.style={dashed,draw=black, postaction={decorate},
        decoration={markings,mark=at position .55 with {\arrow[draw=black]{>}}}},
    scalarbar/.style={dashed,draw=black, postaction={decorate},
        decoration={markings,mark=at position .55 with {\arrow[draw=black]{<}}}},
    scalarnoarrow/.style={dashed,draw=black},
    electron/.style={draw=black, postaction={decorate},
        decoration={markings,mark=at position .55 with {\arrow[draw=black]{>}}}},
	bigvector/.style={decorate, decoration={snake,amplitude=4pt}, draw},
}

\tikzstyle{block} = [draw, rectangle, 
    minimum height=3em, minimum width=6em]
    
\title{
{\Large  \bf Renormalized Schwinger-Dyson functional}
{\vskip 0.5 truecm}} 

\author{{\large Enore Guadagnini$^{\, a}$ and Vittoria Urso$^{\, b}$} \\   
{\normalsize $~$}\\ 
 {\normalsize  $^{ a\, }$Dipartimento di Fisica {\it E. Fermi} dell'Universit\`a di Pisa, and INFN Sezione di Pisa,} \\ {\normalsize   Largo B. Pontecorvo  2, 56127 Pisa, Italy.}
 \\ {\normalsize $^{b\, }$Dipartimento di Matematica e Fisica {\it Ennio De Giorgi}, Universit\`a del Salento,}\\ 
 {\normalsize and IIT di Lecce, Piazza Tancredi  7, 73100 Lecce, Italy. }
 }

\date{}

\begin{document}

\maketitle

\vskip 0.7 truecm 

\begin{abstract}

 We consider the perturbative  renormalization of  the Schwinger-Dyson functional, which is the generating functional of the expectation values of the products of the composite operator given by the field derivative of the action.  
 It is argued that this functional  plays an important role  in the  topological  Chern-Simons  and BF quantum field theories. 
 It is shown that, by means of the  renormalized perturbation theory,   a canonical renormalization procedure for the Schwinger-Dyson functional is obtained. The combinatoric structure of the Feynman diagrams is illustrated in the case of scalar models. For the Chern-Simons and the BF gauge theories, the relationship between the renormalized  Schwinger-Dyson functional and the generating functional of the correlation functions of the gauge fields is produced. \end{abstract}

\vskip 1 truecm

\section{Introduction}

The Schwinger-Dyson equations \cite{D,S} of quantum field theory can be derived   \cite{IZ,PS}  from  the  invariance of the  functional integration under field translations. The  structure of the Schwinger-Dyson equations  is determined by the action functional, which is involved in the computation of  
 the vacuum expectation values of the fields.  
 Let the action  $S[\phi ]$  be a function of a set of fields denoted by $\phi (x) $.  The basic Schwinger-Dyson equation takes the form 
 \be
 \left \langle \frac {\delta S[\phi ]}{\delta \phi (x)} \, \phi(y_1) \phi(y_2) \cdots \phi(y_n)
 \right \rangle =
 i \sum_{j=1}^n \, \delta(x-y_j) \left \langle \phi(y_1) \cdots \phi(y_{j-1}) \phi(y_{j+1}) \cdots \phi(y_n) \right \rangle
  \; ,  
\label{1.1}
\ee
where the vacuum expectation value $\langle {\cal P} [\phi ]  \rangle $ of a field operator ${\cal P} [\phi ] $ is given by 
\be
\left \langle {\cal P} [\phi ]  \right \rangle = \frac{\int D \phi \; e^{iS[\phi ]} \; {\cal P} [\phi ]  }{\int D \phi \; e^{iS[\phi ]}} \; . 
\label{1.2}
\ee
Recently, developments of the Schwinger-Dyson equations  have been applied  in the study of various subjects like, for instance,   the renormalization theory \cite{BD, DKK},  condensed matter investigations \cite{BSF,XYL},  and  bound states and  strong interactions  \cite{CCL,ESW,CR,SAW,PIS,HHT,RBR,RCH,NAT,TAN}.   Standard  Schwinger-Dyson equations have been used also in the case of topological quantum field theories with and without matter \cite{DD,GMS,JTS}. 

We are interested in a particular generalisation of equation (\ref{1.1})  which concerns the computation of  the expectation values of the products of the composite operator $\delta S [\phi ] / \delta \phi (x)$, 
\be
F(x_1,x_2,..., x_n) = \left \langle \frac {\delta S[\phi ]}{\delta \phi (x_1)} \,  \frac {\delta S[\phi ]}{\delta \phi (x_2)} \cdots  \frac {\delta S[\phi ]}{\delta \phi (x_n)}  \right \rangle \; . 
\label{1.3}
\ee 
The generating functional $Z_{SD}[B] $  of the expectation values (\ref{1.3})  is called the Schwinger-Dyson functional and is defined by   
\be
Z_{SD}[B] =  \left \langle e^{i \int dx \, B(x) \, \delta S[\phi ] / \delta \phi (x)  }  \right \rangle  \; ,  
\label{1.4}
\ee
where $B(x)$ denotes a classical source. 

The functional (\ref{1.4}) plays an important role  in the  low-dimensional gauge field theories of topological type, like the Chern-Simons  and BF quantum field theories \cite{EF,MT,E}. 
In these models, the derivative of the action with respect to the components of the connection is proportional to the curvature (plus possible additional contributions which are related to the gauge-fixing lagrangian terms), that  combined  with the topology of 3-manifolds determines the values of the Wilson line observables \cite{EF}.   

In facts, when the gauge structure group of these topological models is abelian, the Schwin\-ger-Dyson functional provides the complete solution for the gauge invariant observables \cite{EF,MT}. 
 
For instance, when the first homology group  \cite{R} $H_1(M)$  of the 3-manifold $M$ is trivial, one can compute \cite{EF} the observables of the abelian Chern-Simons theory defined in $M$ by means of perturbation theory. The action for the connection $A$ is given by $2 \pi k \int A \wedge dA $ and the  variation of the action with respect to the fields is proportional to the curvature  $F_A= dA$. 
By introducing the coupling $\int B \wedge dA $ of the curvature with an external classical source $B = B_\mu (x) dx^\mu $, one finds 
\be
\widetilde Z_{SD} [B] \equiv \left \langle e^{  2 \pi i \int B \wedge dA } \right \rangle = e^{ -i  \, \pi / 2 k \int B\wedge B}  \; , 
\label{1.4b}
\ee   
 that specifies the expectation values of the Wilson lines associated with links in $M$. Quite remarkably, an appropriate generalization \cite{EF} of this procedure  furnishes the solution of the abelian Chern-Simons theory in a generic closed and oriented 3-manifold $M$. Indeed, when the first homology group $H_1(M)$  is not trivial,  for each element of the torsion subgroup \cite{R} of $H_1(M)$ one can introduce a corresponding classical background connection. Then  one needs to take the sum of the Schwinger-Dyson functionals that are computed in the presence of each background connection.  Somehow,  in the functional integration, the values of the curvature  correspond to the local degrees of freedom ---which do not depend on the topology of the manifold $M$---   whereas  the effects of topology  are taken into account by the background connections.    

In the case of the non-abelian $SU(N)$ Chern-Simons theory, the structure of the gauge orbits, which are  associated with the $SU(N)$ connections,    does not admit \cite{SA} a simple description based on the homology group $H_1(M)$.  Yet,  in the characterization of the  local degrees of freedom which are not related with the manifold topology, the non-abelian curvature $F_A = 2 dA + i [A, A]$ appears to play a  fundamental role.  In fact, the value $F_A(x)$ of the curvature in the point $x$ is specified \cite{PS,LA} by  the value of the gauge holonomy associated  an infinitesimal loop centered in $x$, and each infinitesimal loop does not depend on the topology of $M$.  Let us present a  rough sketch of a possible  argument that can be used to make this statement more precise. 

$SU(N)$  gauge connections can be described by one-forms defined in $M$ with values in the $SU(N)$ Lie algebra. The local value $\{ A(x) \} $ of each configuration A can also be specified by the set $ \{ H_\gamma [A] \}$ of the holonomies,  
$$
H_\gamma [A] = {\rm P} e^{i \int_\gamma A} \; , 
$$
which are associated with all the possible closed oriented paths $\{ \gamma \} $ in $M$  with a  given base-point $x_0 \in M$, which represents the  starting/final  point  of each closed path $\gamma$.   This correspondence  is denoted by  
\be
\{ A (x) \} \; \leftrightarrow \;  \{ H_\gamma [A] \}  \; . 
\label{1.4c}
\ee
In turn, the value of the holonomies $H_\gamma [A]$ as a function of the paths can be determined by combining the local values $\{ F_A (x) \} $ of the curvature with smooth deformations of the paths.  In order to illustrate this point, let  
us consider a nontrivial reference path $\gamma_0$, with parametrization $x^\mu (\tau )$ in which  $0 \leq \tau \leq 1$. 
Let the holonomy $\widehat H_{\gamma_0}(s) $ with $0 \leq s \leq 1$ be defined by 
\be
\widehat H_{\gamma_0}(s) =  {\rm P} e^{i \int_0^s d\tau   \, A(\tau ) } \; , 
\label{1.4d}
\ee
in which $A(\tau ) = A_\mu (x (\tau )) (d x^\mu (\tau) / d \tau ) $. Note that $H_{ \gamma_0} [A] =  \widehat H_{\gamma_0}(1) $.  An infinitesimal deformation $\gamma_0 + \delta \gamma $ of the path $\gamma_0$ can be described by the parametrization   $x^\mu (\tau )+ \epsilon^\mu (\tau )$ with $\epsilon^\mu (\tau ) \ll 1 $.  At first order in $ \epsilon^\mu $, one has $H_{\gamma_0 + \delta \gamma} \simeq H_{\gamma_0} + \Delta H_{\gamma_0} $, where the   infinitesimal modification $ \Delta H_{\gamma_0}$ of the holonomy,  
\be
\Delta  H_{\gamma_0} =  i \int_0^1 ds \, \widehat H_{\gamma_0}(s) \, \epsilon^\mu (s) \, \dot x^\nu (s) F_{\mu \nu } (x(s)) \, \widehat H^{-1}_{\gamma_0}(s) \, H_{\gamma_0} [A] \, ,   
\label{1.4e}
\ee
is specified by $\widehat H_{\gamma_0}(s)$ and by the values $\{ F_A (x) \} $ of the curvature. 
If the path $\gamma$ is homotopically equivalent to $\gamma_0$, the value of the associate holonomy $H_\gamma [A]$ is expected to be determined by $ H_{\gamma_0} [A] $ with the help of a set of infinitesimal transformations of the path. As shown in equation (\ref{1.4e}),  the corresponding infinitesimal modifications of the holonomy can be fixed  by means of $\widehat H_{\gamma_0}(s) $ and  the local values  $\{ F_A (x) \} $ of the curvature. If the 3-manifold $M$ is simply connected, any closed path $\gamma$ is homotopic with $\gamma_0$. 
If $M$ is not simply connected,  for each generator $g$ of the fundamental group $\pi_1 (M)$ one can choose a representative path $\gamma_g$ and, in agreement with equation (\ref{1.4d}), one can compute the corresponding holonomy $\widehat H_{\gamma_g} (s)$. For each set  of homotopic paths, the associated holonomies can be determined by combining $\widehat H_{\gamma_g} (s)$ with the local values $\{ F_A (x) \} $ of the curvature.  Thus  relation (\ref{1.4c}) should develop into  
\be
\{ A (x) \} \; \leftrightarrow \;  \{ H_\gamma [A]  \} \; \leftrightarrow \;  \{ \widehat H_{\gamma_1}(s) , \widehat H_{\gamma_2}(s) \, , \dots , \widehat H_{\gamma_g}(s) , \dots \, ; F_A (x) \}  \; . 
\label{1.4f}
\ee
Consequently, in the computation of the functional integration, the sum over the configurations $\{ A (x) \}$ can then be envisaged to be decomposed into a sum over the values $\{ F_A (x) \} $ of the  curvature  and a sum over the values of the  holonomies associated with the  paths which represent the generators of the fundamental group $\pi_1 (M)$. The values of the curvature should correspond to the  ``purely local" degrees of freedom which are independent of the topology of the manifold $M$. Whereas the  values of the  holonomies $\{ \widehat H_{\gamma_1}(s) , \widehat H_{\gamma_2}(s) \, , \dots , \widehat H_{\gamma_g}(s)  \} $ should describe the effects of the nontrivial topology of $M$.  

How to carry out the precise disentanglement of the ``purely local"  degrees of freedom ---described by the curvature---  and the ``topology dependent" degrees of freedom in the functional integration  is an open problem.  In order to investigate this issue  in  the case of the  nonabelian Chern-Simons and BF theories, in the present article we analyze a preliminary question which is related to the perturbative computation of the renormalized generating functional of the vacuum expectation values of the products of the nonabelian curvature $F_A (x)$ in different points of spacetime.  
Indeed, for the topological abelian gauge theories the renormalization is trivial, whereas in the non-abelian case the renormalization task is not trivial. The main purposes of our article is to show how the renormalization of the corresponding Schwinger-Dyson functional  $Z_{SD}[B] $ is canonically determined by the standard renornalization procedure \cite{TC,SOP} for the correlation functions of the gauge connections. 

We demonstrate that, in the Chern-Simons and BF theories, the renormalized Schwinger-Dyson functional  is related with the generating functional $Z[J]$ of the correlation functions of the gauge connections  by some kind of duality transformation. Therefore the standard perturbative procedure called  ``renormalized perturbation theory" \cite{PS} provides a canonical renormalization for $Z_{SD}[B] $. 
Note that we are not interested in the matrix elements 
the composite operator  $\delta S[\phi ] / \delta \phi (x) $
between generic states; this issue can be studied by means of standard techniques \cite{PS,SOP,WW}.  
Motivated by the results of the topological models with an abelian gauge group, we shall concentrate on the vacuum expectation values of products of operators $\delta S[\phi ] / \delta \phi (x) $. In this case, the relationship that we derive between   $Z_{SD}[B] $ and  $Z[J]$ shows that  the standard technique \cite{PS,SOP,WW} for the study of the renormalization properties of the composite operator  $\delta S[\phi ] / \delta \phi (x) $ greatly simplifies.  

Let us remember that the renormalization of the lagrangian field theory models is  expected \cite{C,BS} to be independent of  the global aspects of the manifold that do  not modify  the  short-distance behaviour of the theory.  Therefore, 
since the nonabelian curvature $F_A(x)$ describes degrees of freedom which do not depend on the topology  of the manifold, we shall consider the renormalization properties of  $Z_{SD}[B] $ in flat spacetime.  

The combinatoric structure of the Feynman diagrams ---entering the perturbative computation of $Z_{SD}[B]$--- is illustrated in  the simple case of the field theory models $\phi^3$ and $\phi^4$ in four dimensions in Section~2. 
By means of the Wick contractions \cite{IZ,PS} of the field operators,  we examine the Feynman diagrams which are associated with  the expectation values (\ref{1.3}). We demonstrate that  the short distance behaviour of the products of  the composite operator $\delta S [\phi ] / \delta \phi (x)$ ---that in $d$ dimensional spacetime  has dimension $(d-1)$--- is really determined by the ultraviolet properties  of operators of dimensions $(d-2)$ and/or  $(d-3)$. For instance, when the interacting lagrangian entering the action $S[\phi ]$  is a cubic function of a scalar field $\phi(x)$, $Z_{SD}[B]$ can be related to the ordinary  generating functional $Z[J]$ of correlation functions of the field $\phi (x)$. In general, it turns out  that the connected component of  $Z_{SD}[B]$ is the union of a local functional of  $B(x)$ and a non-local part which is specified by the expectation values of field components   $\phi (x)$ and possibly $\phi^2(x)$.

 Applications and extensions of the results of Section~2 are presented in Section~3, where low dimensional gauge theories of topological type are considered. For the nonabelian $SU(N)$ Chern-Simons model and the $ISU(2)$ BF gauge theory in ${\mathbb R}^3$,  the relationship between the renormalized  Schwinger-Dyson functional and the generating functional of the correlation functions of the gauge fields is produced.    Section~4 contains the conclusions.

\section{Structure of the Feynman diagrams}

The case of a cubic interaction lagrangian is relevant for the topological gauge theories in low dimensions. So, let us first consider the field theory model  which is defined by the action  
\be
S [\phi ] =  \int d^4 x \, \left ( \mezzo \der_\mu \phi \, \der^\mu \phi - \mezzo m^2 \phi^2  + \terzo g \phi^3   \right ) \; , 
\label{2.1}
\ee
where  $\phi (x) $ is a real scalar field and the real parameter $g$ denotes the  coupling constant.  The generating functional $Z[J]$ of the correlation functions of the field $\phi (x)$ is defined by 
\be
Z[J] =  \left \langle  e^{i \int d^4x \, J(x) \,  \phi (x) } \right \rangle   \; .  
\label{2.2}
\ee
The renormalization of $Z[J]$ is obtained by means of the 
standard procedure   denominated  
``renormalized perturbation theory" \cite{PS}. In this scheme,   the  lagrangian parameters assume their renormalized values and, in order to maintain the validity of the  normalization conditions at each order of perturbation theory,  local counterterms are introduced, which  cancel exactly all the contributions to these parameters which are obtained in the  loop expansion.   The normalization conditions for the model defined be the action (\ref{2.1}) concern the values of the mass, of the coupling constant and  the wave function
normalization.  Finally, in order to complete the list of the normalization conditions, one needs to require the absence of a proper vertex which is linear in the field. Let $\Gamma [\varphi ]$ be the effective action which corresponds to the sum of the one-particle-irreducible diagrams with external legs represented by $\varphi (x)$. In agreement with the structure of the lagrangian (\ref{2.1}), the additional normalization condition is given by    $ \left ( \delta \Gamma / \delta \varphi (x) \right ) |_{\varphi =0}  = 0 $.     Note that, in the case of the $\phi^4$ model, the vanishing of the proper vertices which are 
linear and  cubic in powers of the fields is a consequence of the symmetry $ \varphi \rightarrow - \varphi $ which is imposed to the effective action.       In the case of  gauge fields, the analogue of the condition  $ \left ( \delta \Gamma / \delta \varphi (x) \right ) |_{\varphi =0}  = 0 $ is  automatically satisfied. 

Let us now consider the perturbative computation of the mean values (\ref{1.3}). The perturbative expansion \cite{IZ,PS}  of a generic expectation value (\ref{1.2})  can be written as 
\be
\left \langle {\cal P} [\phi ]  \right \rangle = \frac{ \left \langle \, {\cal P} [\phi ] \, e^{i S_I [\phi ]} \, \right \rangle_0 }{\left \langle \,  e^{i S_I [\phi ]  } \, \right \rangle_0   } \; , 
\label{2.3}
\ee
where  $S_I [\phi ]$ denotes the integral of the interaction lagrangian  
\be
S_I [\phi ] =  \terzo g \int d^4x \, \phi^3 (x) \; , 
\label{2.4}
\ee
and the vacuum expectation value of the time-ordered product  of the fields
\be
\left \langle \, {\cal P} [\phi ] \, e^{i S_I [\phi ]} \, \right \rangle_0 \equiv  \langle 0 | \, {\rm T} \left ( {\cal P} [\phi ] \, e^{i S_I [\phi ]} \right ) | 0 \rangle 
\label{2.5}
\ee
corresponds to the sum of the Feynman diagrams which are obtained by means of the Wick contractions \cite{BS}  of the fields.  The set of the connected diagrams is denoted by $\left \langle \, {\cal P} [\phi ] \, e^{i S_I [\phi ]} \, \right \rangle_0^c$.  The Feynman propagator reads  
\be
\langle 0 | \, {\rm T} \left ( \phi (x) \, \phi (y) \right ) | 0 \rangle  = 
\WT{\phi (x) \, \phi}(y) = i \, \langle x | \, \frac{1}{-\der^2 - m^2 + i \epsilon } \, | y \rangle  \; . 
\label{2.6}
\ee
The composite operator $\delta S [\phi ] / \delta \phi (x)$ takes the form  
\be 
\frac{\delta S}{\delta \phi (x)} \equiv E_\phi (x) =   \left [ -\der^2 - m^2 \right ] \phi (x) + g \phi^2 (x)  = \nabla \phi (x) + g \phi^2(x)\; ,  
\label{2.7}
\ee
where we have introduced the simplifying  notation $\nabla \phi (x) \equiv  \left [ -\der^2 - m^2 \right ] \phi (x) $. 
The Schwinger-Dyson functional (\ref{1.4}) can be written as 
$Z_{SD}[B] = \exp (i W_{SD} [B]) $ where $i W_{SD} [B]$ is given by the sum of the connected diagrams 
\be
i W_{SD}[B] = \sum_n \frac{i^n}{n!} \int d^4x_1 \cdots d^4x_n B(x_1) \cdots B(x_n) \, \left \langle  E_\phi (x_1) \cdots E_\phi (x_n) \, e^{i S_I [\phi ]} \, \right \rangle^c_0 \; . 
\label{2.8}
\ee
Note that the overall multiplying factor, which is  given by the sum of the vacuum-to-vacuum diagrams, is not included in the set of the connected diagrams (\ref{2.8}) contributing to $Z_{SD}[B]$.  Indeed,  as a consequence of the normalisation ---shown in equations (\ref{1.2}) and (\ref{2.3})--- of the generating functional (\ref{1.4}), the multiplying factor of the numerator simplifies ---or cancels--- with the same factor of the denominator. 

Let us examine the perturbative evaluation of $\left \langle  E_\phi (x_1) \cdots E_\phi (x_n) \, e^{i S_I [\phi ]} \, \right \rangle^c_0$. In agreement with the Wick Theorem,  let us first consider all the possible  Wick contractions of the operator $\nabla \phi (x)$.    Since $ \left [ -\der^2 - m^2 \right ] \WT{\phi (x) \, \phi}(y) = i \delta^4 (x-y)$, the Wick contraction of $\nabla \phi (x) $ with the fields contained in  $e^{i S_I [\phi ]}$ gives  
\be 
\left \langle  \, \nabla \phi (x)  \, e^{i S_I [\phi ] } \, \right \rangle_0 =  
 i \left \langle   \nabla\WT{\phi (x)  S_I }[\phi ] \, e^{i S_I [\phi ] } \, \right \rangle_0 = \left \langle  (- g \phi^2 (x))   \, e^{i S_I [\phi ] } \, \right \rangle_0 \; , 
\label{2.9}
\ee
and consequently 
\be
 \left \langle E_\phi (x) \, e^{i S_I [\phi ] }   \, \right \rangle^c_0 = \left \langle  \left ( \nabla \phi (x) + g \phi^2(x) \right ) \, e^{i S_I [\phi ] } \, \right \rangle^c_0 = 0 \; . 
 \label{2.10}
\ee
Let us now consider $ \left \langle E_\phi (x_1) E_\phi (x_2) \, e^{i S_I [\phi ] }   \, \right \rangle^c_0 $. Because of equation (\ref{2.9}), the contraction of $\nabla \phi (x_1)$ with the fields contained in $e^{i S_I [\phi ]}$ gives a vanishing result as a consequence of the sum with the term $g \phi^2 (x_1)$, as 
 shown in equation (\ref{2.10}) for the case of $ \left \langle E_\phi (x) \, e^{i S_I [\phi ] }   \, \right \rangle^c_0 $.  
 So we must consider the contraction of $\nabla \phi (x_1)$ with $E_\phi (x_2)$, which  produces   
\be
 {\nabla\WT{ \phi (x_1)  E}}_\phi (x_2) =  i  \left [ -\der^2 - m^2 \right ] \delta^4 (x_1- x_2 ) + 2 i g \,  \phi (x_2)  \, \delta^4 (x_1- x_2 )    \; . 
\label{2.11}
\ee
Thus one finds   
\bea
\left \langle E_\phi (x_1) E_\phi (x_2) \, e^{i S_I [\phi ] }   \, \right \rangle^c_0 &=& i \left [ -\der^2 - m^2 \right ] \, \delta^4 (x_1-x_2) \nonumber \\ 
&& \qquad + 2 i g \delta^4 (x_1 - x_2) \left \langle  \phi (x_2) \, e^{i S_I [\phi ]} \right \rangle_0^c \; . 
 \label{2.12}
\eea
The normalization condition $ \left ( \delta \Gamma / \delta \varphi (x) \right ) |_{\varphi =0}  = 0 $ on the  absence of the tadpole implies  \newline  $\left \langle  \phi (x_2) \, e^{i S_I [\phi ]} \right \rangle_0^c = 0$. Therefore 
\be
\left \langle E_\phi (x_1) E_\phi (x_2) \, e^{i S_I [\phi ] }   \, \right \rangle^c_0 = i \left [ -\der^2 - m^2 \right ] \, \delta^4 (x_1-x_2)  \; . 
 \label{2.13}
\ee
The same arguments illustrated above give 
\be
\left \langle E_\phi (x_1) E_\phi (x_2) E_\phi (x_3) \, e^{i S_I [\phi ] }   \, \right \rangle^c_0 = - 4  \, g\,  \delta^4 (x_1-x_2)  \, \delta^4 (x_2 - x_3)\; . 
 \label{2.14}
\ee
The structure of the diagrams associated with $\left \langle E_\phi (x_1) E_\phi (x_2) \cdots  E_\phi (x_n) \, e^{i S_I [\phi ] }   \, \right \rangle^c_0  $, for generic  $n \geq 4$, can be obtained by first considering  all the Wick contractions of the field operators of the type $\nabla \phi $. 
The combinatoric of these contractions can easily be obtained by taking into account the symmetric role of the operators $E_\phi (x_1) E_\phi (x_2) \cdots  E_\phi (x_n)$ in the computation of $W_{SD} [B]$, as shown in equation (\ref{2.8}). For the connected diagrams, we find  
\be
\left \langle E_\phi (x_1) E_\phi (x_2) \cdots  E_\phi (x_n) \, e^{i S_I [\phi ] }   \, \right \rangle^c_0 = 0 \quad , \quad \hbox{ if $n$ is odd} \; , 
\label{2.15}
\ee
and 
\bea
\left \langle E_\phi (x_1) E_\phi (x_2) \cdots  E_\phi (x_n) \, e^{i S_I [\phi ] }   \, \right \rangle^c_0 &=&   \left \langle \phi (x_2) \, \phi (x_4) \cdots  \phi (x_n) \, e^{i S_I [\phi ] }   \, \right \rangle^c_0 \times \nonumber \\ 
&&{\hskip - 6 cm} \times (n-1)!! \, (2 i g )^{n/2} \, \delta^4 (x_1 - x_2) \cdots \delta^4 (x_{n-1} - x_n) 
 \quad , \quad \hbox{ if $n$ is even} \; .  
\label{2.16}
\eea
Equations (\ref{2.15}) and (\ref{2.16}) show that the expectation value  $\left \langle E_\phi (x_1) E_\phi (x_2) \cdots  E_\phi (x_n) \, e^{i S_I [\phi ] }   \, \right \rangle^c_0  $ is completely specified by the expectation value   $\left \langle \phi (x_2) \, \phi (x_4) \cdots  \phi (x_n) \, e^{i S_I [\phi ] }   \, \right \rangle^c_0$. 
Therefore the standard renormalization procedure for the correlation functions of the field $\phi (x)$ canonically defines the renormalization for $\left \langle E_\phi (x_1) E_\phi (x_2) \cdots  E_\phi (x_n) \, e^{i S_I [\phi ] }   \, \right \rangle^c_0  $. 
Equation (\ref{2.10}) implies 
\be
i \int d^4x \, B(x)  \, \left \langle  E_\phi (x) \, e^{i S_I [\phi ]} \, \right \rangle^c_0 = 0 \; ,
\label{2.16a}
\ee
The contact terms (\ref{2.13}) and (\ref{2.14}) give origin to a local contribution to $W_{SD}[B]$. In particular,  we find 
\be
 \frac{i^2}{2!} \int d^4x_1 d^4x_2 \, B(x_1) B(x_2) \, \left \langle  E_\phi (x_1)  E_\phi (x_2) \, e^{i S_I [\phi ]} \, \right \rangle^c_0 = i \int d^4x \, \left \{ - \mezzo  \der_\mu B \der^\mu B + \mezzo m^2 B^2  \right \}\; , 
\label{2.16b}
\ee
\be
\frac{i^3}{3!} \int d^4x_1 d^4x_2 d^4x_3 \, B(x_1) B(x_2) B(x_n) \, \left \langle  E_\phi (x_1) E_\phi (x_2) E_\phi (x_3) \, e^{i S_I [\phi ]} \, \right \rangle^c_0 =  i \int d^4 x   \, \left [  \dueterzi g B^3  \right ] \; .
\label{2.16c} 
\ee
The expectation values   $\left \langle \phi (x_2) \, \phi (x_4) \cdots  \phi (x_n) \, e^{i S_I [\phi ] }   \, \right \rangle^c_0$, with $n \geq 4$,   correspond to non-local amplitudes. By collecting all the results  on $\left \langle E_\phi (x_1) E_\phi (x_2) \cdots  E_\phi (x_n) \, e^{i S_I [\phi ] }   \, \right \rangle^c_0  $ with $n\geq 4$,  we get 
\bea
&&\sum_{n=4}^\infty \frac{i^n}{n!} \int d^4x_1 \cdots d^4x_n B(x_1) \cdots B(x_n) \, \left \langle  E_\phi (x_1) \cdots E_\phi (x_n) \, e^{i S_I [\phi ]} \, \right \rangle^c_0 = \nonumber \\ 
&& \quad = 
\sum_{p=2}^\infty \frac{i^{(2p)}  (2p -1)!! \, (2 i g )^{p}  }{(2p)!} \int d^4x_1 \cdots d^4x_{2p} B(x_1) \cdots B(x_{2p}) \, \times \nonumber \\ 
&& \qquad \qquad \qquad \times 
 \left \langle \phi (x_2) \, \phi (x_4) \cdots  \phi (x_{2p}) \, e^{i S_I [\phi ] }   \, \right \rangle^c_0
\delta^4 (x_1 - x_2) \cdots \delta^4 (x_{2p-1} - x_{2p}) =  \nonumber \\ 
&& \qquad = 
\sum_{p=2}^\infty \frac{(- i g )^{p}  }{p!} \int d^4x_1 \cdots d^4x_{p} B^2(x_1) \cdots B^2(x_p)  
 \left \langle \phi (x_1) \, \phi (x_2) \cdots  \phi (x_p) \, e^{i S_I [\phi ] }   \, \right \rangle^c_0 = \nonumber \\ 
 && \qquad = 
 \left \langle e^{i \int d^4x (-g) B^2 (x) \phi (x)} \, e^{i S_I [\phi ] }   \> \right \rangle^c_0 = Z \left [ \left ( J= - g B^2 \right ) \right ]  \; . 
\label{2.16d}
\eea
The sum of the  contributions (\ref{2.16a})-(\ref{2.16d}) shows    that the renormalized Schwinger-Dyson functional for the $\phi^3$ model satisfies 
\be
Z_{SD}[B ] = e^{iR[ B ]} \, Z[\widetilde J[B]  ]\; , 
\label{2.17}
\ee
where
\be
R[B] =  \int d^3x \, \left \{ - \mezzo  \der_\mu B \der^\mu B + \mezzo m^2 B^2  + \dueterzi g B^3  \right \}\; , 
\label{2.18}
\ee
and
\be
\widetilde J[B](x) = - g B^2(x)\; . 
\label{2.19}
\ee
The expectation values (\ref{2.13}) and (\ref{2.14}) determine the local functional $R[B]$ of equation (\ref{2.18}). 
As shown in equation (\ref{2.17}), the renormalization of $Z_{SD}[B ] $ is specified by the standard renormalization of $Z[J ]$. 

The structure of the results for the $\phi^3$ model admits  appropriate  generalizations which depend on the form of the lagrangian of each field theory.  Let us consider for instance the  $\phi^4$ model which is specified by the action 
\be
S [\phi ] =  \int d^4 x \, \left ( \mezzo \der_\mu \phi \, \der^\mu \phi - \mezzo m^2 \phi^2  - \quarto g \phi^4   \right ) \; .  
\label{2.20}
\ee
One has 
\be 
\frac{\delta S}{\delta \phi (x)} =   \left [ -\der^2 - m^2 \right ] \phi (x) - g \phi^3 (x)  \; .  
\label{2.21}
\ee
The perturbative expansion of the Schwinger-Dyson functional $Z_{SD} [B]$ can be  examined  by means of the method described above. We find 
\be
Z_{SD}[B ] = e^{iQ[ B ]}  \left \langle 
\exp \left \{ i    \int d^4 x \, \left [   -2 g B^3(x) \, \phi (x) + \tremezzi g B^2(x) \, \phi^2 (x) \right ]  \right \}
\right \rangle 
\; , 
\label{2.22}
\ee
in which 
\be
Q[B] = \int d^4 x \left [   - \mezzo \der_\mu B(x) \der^\mu B(x) + \mezzo m^2 B^2 (x) + \trequarti g B^4 (x)  \right ] \; . 
\label{2.23}
\ee
In this case,  $Z_{SD}[B ] $ is related to the expectation value of a  term in which, in addition to a  coupling with the field operator $\phi (x)$,   a coupling with the operator $\phi^2(x)$ is also present.  Note that the short distance behaviour of the composite operator $\phi^2 (x)$ is taken into account by the standard renormalization of the generating functional $Z[J]$ of the correlation functions  because $\phi^2 (x)$ has canonical dimension $2 $. For instance, the one-loop correlation $\langle \phi^2 (x) \phi^2 (y) \rangle $ is described by the diagram of Figure~1 (with removed external legs), which enters   the ordinary renormalization of the $\phi^4$ theory. 
Thus, for this model also, the renormalization of $Z_{SD}[B ] $ can be specified by the standard renormalization procedure. 

\vskip 0.6 truecm 
\centerline {
\begin{tikzpicture}[line width=1.1 pt, scale=0.8]
%
\draw (0.8,0) ellipse (0.8 and 0.5);
\draw(-1,1) -- (0,0); 
\draw(-1,-1) -- (0,0); 
\draw(2.6,1) -- (1.6,0); 
\draw(2.6,-1) -- (1.6,0); 
\draw[fill=black] (0,0) circle (0.03cm);
\draw[fill=black] (1.6,0) circle (0.03cm); 
 \end{tikzpicture}}
\vskip 0.4 truecm 
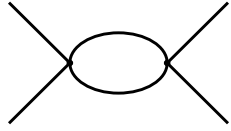
\centerline {Figure 1.  One loop diagram of the $\phi^4$ model  contributing to $\langle \phi^2 (x) \phi^2 (y) \rangle $.}
\vskip 0.5 truecm

As these examples have shown, the connected component $W_{SD} [B]$ of the renormalized Schwinger-Dyson functional contains a local part, expressions (\ref{2.18}) and (\ref{2.23}), which is determined by the form of the lagrangian  of each specific model, and a non-local contribution which is related to the vacuum expectation values  of field operators of dimension 1 or 2, {\em i.e.} the fields $\phi (x)$ and $\phi^2 (x)$.   This universal feature can be used to specify the renormalized values of the parameters of each model by the introduction of appropriate normalization conditions written in terms of the operator $\left [ \delta S / \delta \phi (x) \right ] $. Equivalently, some of the vacuum expectations values of the products of $\left [ \delta S / \delta \phi (x) \right ] $ are uniquely specified and do not receive perturbative changes. For example, in the $\phi^3$ model, relations (\ref{2.13}) and (\ref{2.14}) are exact and are not modified by loop corrections. 

\section{Topological models}

In this section we consider  gauge theories of topological type in ${\mathbb R}^3$. The action of the  $SU(N)$ quantum Chern-Simons theory \cite{W,GMM,E} in the Landau gauge is given by 
\bea
S  &=& {k \over 4 \pi} \int d^3x \,\Big \{  \epsilon^{\mu \nu \tau } \left [ \mezzo  A^a_\mu \der_\nu A^a_\tau - \sesto \, f^{abc} A^a_\mu A^b_\nu A^c_\tau \right ] \nonumber \\ 
&& \qquad  \qquad - M^a\der^\mu A^a_\mu + \der^\mu {\overline c}^a   \left ( \der_\mu c^a - f^{abc} A^b_\mu c^c \right ) \Big \} \; , 
\label{3.1}
\eea
and then 
\be 
\frac {\delta S}{\delta A^a_\mu (x)} =   \left ( { k \over 4 \pi } \right ) \left [ \epsilon^{\mu \nu \tau }\left ( \der_\nu A_\tau^a (x) - \mezzo \, f^{abc}  A^b_\nu (x) A^c_\tau (x) \right ) + \der^\mu M^a (x) + \der^\mu {\overline c}^b (x) f^{abd}  c^d (x) \right ] \; . 
\label{3.2}
\ee
The Schwinger-Dyson functional $Z_{SD} [L^a_\mu ]$ is defined by 
\be
 Z_{SD}[L^a_\mu ] =   \left \langle \exp \left ( i \int d^3x \, L^a_\mu(x)  \left [ \delta S  / \delta A_\mu^a(x) \right ] \right ) \right \rangle \; , 
 \label{3.3}
 \ee
and the generating functional $Z[J^a_\mu ]$ of the correlation functions for the gauge field $A^a_\mu (x)$ is given by 
\be
Z[J^a_\mu ] = \left \langle \exp \left ( i \int d^3x \, J^{ a \mu } (x)  A_\mu^a(x)  \right ) \right \rangle \; . 
 \label{3.4}
\ee
In order to examine the diagrams entering the vacuum expectation values of  the product of fields $\delta S / \delta A^a_\mu (x_1) \cdots \delta S / \delta A^b_\nu (x_n) $, one needs to use the following relationship between the propagators of the fields 
\be
\epsilon^{\mu \sigma \lambda} \der_\sigma  \WT{A^a_\lambda (x)  A} \! \null^b_\nu (y) + \der^\mu \! \WT{M^a (x) A}\! \null^b_\nu (y) = i \left ( \frac{4 \pi}{k}\right ) \delta^{ab}\, \delta^\mu_\nu \, \delta^3 (x-y) \; , 
\label{3.5}
\ee
which can be derived  from the action (\ref{3.1}), or it can be checked directly  by means of the expressions    
\be
\WT{A^a_\mu (x) A}\! \null^b_\nu(y) = \delta^{ab} \left ( \frac{4 \pi}{k}\right ) \int {d^3p \over (2 \pi )^3} \, e^{i p(x-y)} \, \epsilon_{\mu \nu \lambda} {p^\lambda \over p^2}  \; ,
\label{3.6}
\ee
\be 
\WT{A^a_\mu (x)M}\! \null^b(y) = - \delta^{ab} \left ( \frac{4 \pi}{k}\right ) \int {d^3p \over (2 \pi )^3} \, e^{i p(x-y)} \, {p_\mu \over p^2} \; . 
\label{3.7}
\ee 
The presence of the operator $ \der^\mu {\overline c}^b (x) f^{abd}  c^d (x)  $ in the equation (\ref{3.2}) does not modify  the perturbative relations between the expectation values of the operators  $\delta S / \delta A^a_\mu (x_1) \cdots \delta S / \delta A^b_\nu (x_n) $ ---that have been derived in Section~2 by using  the Wick contractions of the fields---  because this additional term has no contractions with the gauge fields $A^a_\mu (x)$ and the auxiliary field $M^a(x)$. Thus, by means of  the arguments presented in Section~2 , one gets 
\be
Z_{SD}[L^a_\mu ] = e^{iF[L^a_\mu]} \, Z[\, \widehat J[ L]^a_ \mu \, ]\; , 
\label{3.8}
\ee
where
\be
F[L^a_\mu] = - {k \over 4 \pi} \int d^3x \, \epsilon^{\mu \nu \tau } \left \{ \mezzo L^a_\mu \der_\nu L^a_\tau + \terzo f^{abc} L^a_\mu L^b_\nu L^c_\tau \right \}\; , 
\label{3.9}
\ee
and
\be
\widehat J[L]^a_\mu(x) = {k \over 8 \pi } \epsilon^{\mu \nu \tau} \, f^{abc} L_\nu^b(x) L^c_\tau (x)\; . 
\label{3.10}
\ee
In addition to the  local component which is described the function $ F[L^a_\mu]$ of equation  (\ref{3.9}), the connected
component of  $Z_{SD}[L^a_\mu ] $ contains powers of $L^a_\mu $  greater than (or equal to)  four.  In particular, for  each connected diagram $D$ contributing to $Z[J^a_\mu ]$ there is a corresponding connected diagram $ D^\prime $ contributing to $Z_{SD}[L^a_\mu ]$. In agreement with equation (\ref{3.8}),   $ D^\prime $ can be obtained from $D$ by the  introduction of the ``effective classical vertex" (\ref{3.10}) in each external leg of $D$, as depicted in Figure~2.  Clearly, the renormalization of the amplitude associated with each diagram $D$  defines a canonical renormalization of the amplitude corresponding to $D^\prime$. 
So, equation (\ref{3.8}) provides a definition of the renormalized Schwinger-Dyson functional $Z_{SD} [L^a_\mu ]$. 

\vskip 0.4 truecm 
\centerline {
\begin{tikzpicture}[line width=1.1 pt, scale=0.6]
\draw [ pattern=north east lines ] (2,1) circle (1);
\draw(0,1) -- (1,1);
\draw(3,1) -- (4,1);
\draw(2,2) -- (2,3);
\node at (2,-1.5) {{$D$}};
\draw [ pattern=north east lines ]  (11,1) circle (1);
\draw(9,1) -- (10,1);
\draw(12,1) -- (13,1);
\draw(11,2) -- (11,3);
\node at (11,-1.5) {{$D^\prime $}};
\draw[dashed](8,0) -- (9,1);  
\draw[dashed](8,2) -- (9,1);
\draw[dashed](10,4) -- (11,3);
\draw[dashed](12,4) -- (11,3);
\draw[dashed](13,1) -- (14,0);
\draw[dashed](13,1) -- (14,2);
 \end{tikzpicture}
 }
\vskip 0.6 truecm 
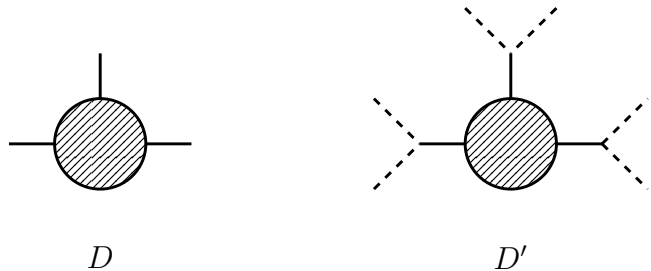
\centerline {{Figure~2.} ~Diagram $D$ and the associated diagram $D^\prime$ contributing to $Z_{SD}$.}
\vskip 0.5 truecm

Equation (\ref{3.8}) also specifies the leading term of the operator product expansion  \cite{PS}
\be
\left [ \delta S / \delta A^a_\mu (0) \right ]  \, \left [ \delta S / \delta A^b_\nu (x) \right ]  \rightarrow \sum_n C_n(x) \, {\cal O}_n  \; . 
\label{3.11}
\ee
Indeed the value of the coefficient function $C_1(x)$ for the identity operator ${\cal O}_1 = 1$ is determined by expression (\ref{3.9})
\be
C_1(x) = \left (   \frac{k}{4 \pi } \right )  \delta^{ab} \epsilon^{\mu \nu \lambda } \der_\lambda \delta^3 (x) \; ,  
\label{3.12}
\ee
and it does  not receive perturbative corrections. 

The Schwinger-Dyson functional for the abelian Chern-Simons theory can be obtained in the  $f^{abc} \rightarrow 0 $ limit.  In this case, equation (\ref{3.8}) becomes  
\be
Z_{SD}[L_\mu ] = \exp \left \{  -i  {k \over 4 \pi}   \int d^3x \, \epsilon^{\mu \nu \tau } \, \mezzo L_\mu \der_\nu L_\tau  \right \}   \; . 
\label{3.13}
\ee
When $L_\mu (x)$ coincides with the de~Rham-Federer current \cite{DR,HF,BGST,EFS} associated with a link $\cal L$ (with support on a Seifert surface \cite{R} which bounds the link $\cal L\, $), expression (\ref{3.13}) represents precisely the exponent of the  linking matrix   corresponding to $\cal L$. For this reason, in the abelian Chern-Simons theory the Schwinger-Dyson functional provides the solution  \cite{EF} for the link observables. 

Finally, let us consider the $ISU(2)$ BF theory \cite{HO,KR,MP,BT,GR}  in ${\mathbb R}^3$, with  fields $A^a_\mu (x)$ and $B^a_\mu (s)$ and gauge-fixed  action in the Landau gauge 
\bea
S &=& \int d^3 x \, \epsilon^{\mu \nu \lambda}  \left \{ \mezzo  \, B^a_\mu F^a_{\nu \lambda} (A) + g  \left [ \mezzo A^a_\mu \der_\nu A^a_\lambda
 - \sesto \epsilon^{abc }\, A^a_\mu A^b_\nu A^c_\lambda \right ] \right \}  \nonumber \\ 
   &&\qquad +  \int d^3x \Bigl \{ M^a \der^\mu A^a_\mu + N^a \der^\mu B^a_\mu + \der^\mu  {\overline c}^a  (\der_\mu c^a - \epsilon^{abc} A_\mu^b c^c ) \nonumber \\ 
&& {\hskip 1.8 cm} + \der^\mu  {\overline \xi}^a  (\der_\mu \xi^a - \epsilon^{abc} A_\mu^b \xi^c
- \epsilon^{abc} B_\mu^b c^c) \Bigr \} \; ,  
 \label{3.14}
 \eea
where the real parameter $g$ denotes a coupling constant and 
\be
F^a_{\mu \nu }(A) = \der_\mu A_\nu^a - \der_\nu A_\mu^a  - \epsilon^{abc} A^b_\mu  A^c_\nu \; . 
\label{3.15}
\ee
The non-vanishing propagators for the components of the connection and the auxiliary fields are given by \cite{GR}
\bea
\WT{A^a_\mu (x) B}\! \null^b_\nu(y) &=& \delta^{ab} \int {d^3k \over (2 \pi )^3} \, e^{i k(x-y)} \, \epsilon_{\mu \nu \lambda} {k^\lambda \over k^2}   \; , \nonumber \\ 
\WT{B^a_\mu (x) B}\! \null^b_\nu(y) &=& - g \, \delta^{ab} \int {d^3k \over (2 \pi )^3} \, e^{i k(x-y)} \, \epsilon_{\mu \nu \lambda} {k^\lambda \over k^2}  \; , 
\label{3.16}
\eea
and
\bea
\WT{A^a_\mu (x)M}\! \null^b(y) &=& \delta^{ab} \int {d^3k \over (2 \pi )^3} \, e^{i k(x-y)} \, {k_\mu \over k^2} \; , \nonumber \\ 
\WT{B^a_\mu (x)N}\! \null^b(y) &=&  \delta^{ab} \int {d^3k \over (2 \pi )^3} \, e^{i k(x-y)} \, {k_\mu \over k^2}   \; . 
\label{3.17}
\eea
In this model one has 
\be
\frac{\delta S}{\delta A^a_\mu (x)}  = \epsilon^{\mu \nu \lambda } \left [ \left ( \der_\nu B^a_\lambda  - \epsilon^{abc} A^b_\nu B_c^c  \right )   + \gmezzi F^a_{\nu \lambda} (A) \right ]  - \der^\mu M^a - \epsilon^{abc} \left [ \der^\mu  {\overline c}^a  c^c  -  \der^\mu  {\overline \xi}^a    \xi^c \right ] ,
\label{3.18}
\ee
and 
\be
\frac{\delta S}{\delta B^a_\mu (x)} = \mezzo \epsilon^{\mu \nu \lambda } F^a_{\nu \lambda } (A) - \der^\mu N^a - \epsilon^{abc} \der^\mu  {\overline \xi}^a c^c \; . 
\label{3.19}
\ee
The Schwinger-Dyson functional $Z_{SD} [L^a_\mu , H^a_\mu ]$ is defined by 
\be
 Z_{SD}[L^a_\mu , H^a_\mu  ] =   \left \langle e^{i   \int d^3x \left [  L^a_\mu  \left ( \delta S  / \delta A_\mu^a \right ) +  H^a_\mu  \left ( \delta S  / \delta B_\mu^a \right ) \right ] } \right \rangle \; , 
 \label{3.20}
 \ee
where $L^a_\mu (x)$ and $H^a_\mu (x)$ are classical sources.  The generating functional $Z[J^a_\mu , K^a_\mu   ]$ of the correlation functions for the gauge field $A^a_\mu (x)$ and $B_\mu^a(x)$  is given by 
\be
Z[J^a_\mu , k^a_\mu  ] = \left \langle e^{ i \int d^3x \left [  J^a_ \mu    A_\mu^a + K^a_\mu B^a_\mu   \right ]} \right \rangle \; . 
 \label{3.21}
\ee
With several field components, the construction and the sum of the Feynman diagrams becomes rather laborious. We get 
\be
 Z_{SD}[L^a_\mu , H^a_\mu  ] =  e^{i G[ L^a_\mu , H^a_\mu ]} \, Z[\widetilde J^a_\mu , \widetilde K^a_\mu  ] \; , 
 \label{3.22}
 \ee
in which 
\be
G[ L^a_\mu , H^a_\mu ] = - \int d^3x \, \epsilon^{\mu \nu \lambda } \Bigl \{ H^a_\mu \der_\nu L^a_\lambda + \epsilon^{abc} H^a_\mu L^b_\nu L^c_\lambda  
+ \gmezzi L^a_\mu \der_\nu L^a_\lambda + \gterzi \epsilon^{abc} L^a_\mu L^b_\nu L^c_\lambda \Bigr \} , 
\label{3.23}
\ee
and 
\be
\widetilde J^a_\mu = \epsilon^{\mu \nu \lambda} \epsilon^{abc}   \left [  H^b_\nu L^c_\lambda + \gmezzi L^b_\nu L^c_\lambda  \  \right ] \qquad  , \qquad  \widetilde K^a_\mu = \mezzo \epsilon^{\mu \nu \lambda} \epsilon^{abc} L^b_\nu L^c_\lambda \; . 
\label{3.24}
\ee

In addition to the contact terms, which are specified by the local functional $G[ L^a_\mu , H^a_\mu ] $, in the BF theory there are additional expectation values that can be displayed in their exact form. For instance, since the vacuum polarization vanishes, from equation (\ref{3.22}) one derives  the following relation for the connected mean value   
\bea 
&& \int d^3x d^3y d^3z d^3t \,  L^a_\mu (x) L^b_\nu (y) L^c_\lambda (z) H^d_\rho (t)    \left \langle   \frac{\delta S}{A^a_\mu (x)}   \frac{\delta S}{A^b_\nu (y)}  \frac{\delta S}{A^c_\lambda (z)} \frac{\delta S}{B^d_\rho (t)}   \right  \rangle^c = \nonumber \\ 
&& \; \;  = -i \frac {3}{4 \pi } \int d^3x \, d^3y \, \frac{(x-y)^\lambda}{|x-y|^3} \epsilon_{\lambda \mu \nu  } \epsilon^{\mu \rho \tau } \epsilon^{\nu \sigma \xi } \epsilon^{abc} \epsilon^{ade} \, H^b_\rho (x) L^c_\tau (x) L^d_\sigma (y) L^e_\xi (y) \; . 
\label{3.25}
\eea 
Actually in both the Chern-Simons and the BF theories, the connected components of the Schwinger-Dyson functional  
containing up to six powers of the external classical sources are exhibited in closed form because of the vanishing of the loop  corrections to the two-points and three-points correlation functions of the gauge fields \cite{SOP,GMM,GR}. 

\section{Conclusions}

For  renormalizable quantum field theories  we have shown that,   in the perturbative computation of the corresponding  Schwinger-Dyson functional  $Z_{SD}$, the use of standard renormalized perturbation theory  provides a canonical renormalization procedure for  $Z_{SD}$. In facts, the short distance behaviour of the products of  the composite operator $\delta S [\phi ] / \delta \phi (x)$ turns out to be  determined by the ultraviolet properties  of the field operators of dimensions 1 (and possibly 2).  The explicit combinatoric of the Wick contractions of the field operators  and the resulting structure of the Feynman diagrams have  been illustrated in the simple cases of the  $\phi^3$ and $\phi^4$ models. We have shown that the connected component of  $Z_{SD}$ is the union of a local functional of the classical source and a non-local part which is specified by the expectation values of field components. The arguments that have been presented in these  scalar models naturally extend to a generic theory.  

In order to study possible applications of the Schwinger-Dyson functional in  gauge field theories of topological type, for the non-abelian Chern-Simons and  BF gauge theories,  the relationship between the renormalized  Schwinger-Dyson functional and the generating functional of the correlation functions of the gauge fields has been derived.   In these cases, the vanishing of the loop  corrections  for the two-points and three-points correlation  functions implies that the connected components of the renormalized  Schwinger-Dyson functional containing up to six powers of the external classical sources have been produced in closed form. 
In these topological models, the derivative of the action with respect of the gauge fields is proportional to the curvature of the connection (plus gauge-fixing contributions). So   
relations (\ref{3.8}) and (\ref{3.22})  could possibly be used for the introduction of appropriate field variables ---similar to  
the local gauge invariant variables decomposition \cite{GS}--- 
 which simplify the functional integration when the theory is defined in topological non-trivial manifolds. 

 \vskip 1.4 truecm 

\bibliographystyle{amsalpha}

\end{document}